\documentclass[aip,amsmath,amssymb,reprint,superscriptaddress]{revtex4-1}

\usepackage{graphicx}
\usepackage{calligra}
\usepackage{upgreek}
\usepackage{color}
\usepackage{inputenc,caption}
\usepackage{stackrel}
\usepackage{subcaption}
\usepackage{placeins}
\usepackage{amsmath}
\usepackage{soul}
\usepackage{hyperref}

\DeclareMathAlphabet{\mathpzc}{OT1}{pzc}{m}{it}

\newcommand{\be}{\begin{equation}}
\newcommand{\ee}{\end{equation}}
\newcommand{\bea}{\begin{eqnarray}}
\newcommand{\eea}{\end{eqnarray}}
\newcommand{\lb}{\label}

\newcommand{\bx}{{\bf x}}

\newcommand{\bsigma}{{\mbox{\boldmath $\sigma$}}}

\newcommand{\grad}{{\mbox{\boldmath $\nabla$}}}

\begin{document}

\setlength{\abovedisplayskip}{8pt}
\setlength{\belowdisplayskip}{8pt}

\title[]{Emergence of long-range non-equilibrium correlations in free liquid diffusion}

\author{Marco Bussoletti}
\affiliation{Department of Mechanical and Aerospace Engineering, Sapienza University of Rome, via Eudossiana 18, 00184 Rome, Italy}
\author{Mirko Gallo}
\affiliation{Department of Mechanical and Aerospace Engineering, Sapienza University of Rome, via Eudossiana 18, 00184 Rome, Italy}
\author{Amir Jafari}
\affiliation{Department of Applied Mathematics and Statistics, The Johns Hopkins University, Baltimore, MD, USA, 21218}
\affiliation{Current address: 11 Sheldon Square, London, United Kingdom, W26DQ} 
\author{Gregory L. Eyink}
\affiliation{Department of Applied Mathematics and Statistics, The Johns Hopkins University, Baltimore, MD, USA, 21218}
\affiliation{Department of Physics, The Johns Hopkins University, Baltimore, MD, USA, 21218}

\begin{abstract}
It is experimentally well-established that non-equilibrium long-range correlations 
of concentration fluctuations appear in free diffusion of a solute in a solvent, but 
it remains unknown how such correlations are established dynamically. We address this problem in a 
model of Donev, Fai \& Vanden-Eijnden (DFV), obtained from the high-Schmidt limit of the 
Landau-Lifschitz fluctuating hydrodynamic equations for a binary mixture. 
We consider an initial planar interface of the mean concentration field in an infinite 
space domain, idealizing prior experiments. Using methods borrowed from turbulence theory, 
we show both analytically and numerically that a 
quasi-steady regime with self-similar time decay of concentration correlations appears 
at long time. In addition to the expected ``giant concentration fluctuations'' with 
correlations $\propto r$ for $r\lesssim L(t)=(Dt)^{1/2},$ with diffusivity $D,$ 
a new regime with spatial decay 
$\propto 1/r$ appears for $r\gtrsim L(t).$ The quasi-steady regime arises from 
an initial stage of transient growth $\propto t,$ confirming the prediction of DFV
for $r\gtrsim L(t)$ and discovering an analogous result for $r\lesssim L(t).$ Our results 
give new insight into the emergence of non-equilibrium long-range correlations and provide 
novel predictions that may be investigated experimentally. 
\end{abstract}

\maketitle

\section{\label{sec:intro}Introduction}

One of the important discoveries in statistical physics of the late 20th 
century was generic spatial long-range correlations
in non-equilibrium steady-states, first predicted theoretically 
\cite{kirkpatrick1979kinetic,kirkpatrick1980hydrodynamic,kirkpatrick1982light,ronis1979statistical,ronis1980statistical}
and subsequently confirmed in laboratory experiments \cite{dezarate2006hydrodynamic}.  
Such non-equilibrium long-range correlations have been observed experimentally even in 
transient regimes, notably in free diffusion of an initial blob of solute in a liquid solution 
\cite{vailati1997giant}. These so-called ``giant concentration fluctuations'' (GCF's) were successfully 
explained by linearized fluctuating hydrodynamics \cite{vailati1998nonequilibrium}, 
patterned after the earlier theoretical work on non-equilibrium steady-states \cite{ronis1980statistical}. 
In low-gravity experiments without quenching by bouyancy effects these GCF's have been 
observed to propagate even up to the system size, on order of centimeters 
\cite{croccolo2016shadowgraph}. Many questions remain open in this area, both theoretically 
and experimentally, and are the subject of intensive on-going investigations \cite{vailati2024perspective}.
One general fundamental question is the dynamical emergence of such long-range non-equilibrium 
correlations \cite{doyon2023emergence}, which is the subject of the present paper for the 
particular case of concentration fluctuations in free liquid diffusion.

Our work is based on the remarkable theory of Donev, Fai \& vanden-Eijnden (DFV) \cite{donev2014reversible},
which revealed a close analogy between the emergence of GCF's in free liquid 
diffusion and the cascade of a passive scalar advected by turbulent velocity field. More precisely, DFV 
considered nonlinear fluctuating hydrodynamics for the concentration field advected by the thermal 
fluctuating velocity field in the high Schmidt-number limit, $Sc=\nu/D\gg 1, $ a condition which holds rather generally 
for liquid mixtures. Because the velocity dynamics associated with kinematic viscosity $\nu$  is 
very fast in this limit compared with the dynamics of the concentration associated to diffusivity $D,$
the fluctuating hydrodynamic equation for the latter reduces to a version of the {\it Kraichnan model} 
of a turbulent scalar, in which the advecting velocity is Gaussian 
and white-in-time \cite{kraichnan1968small,falkovich2001particles}. Building upon earlier more heuristic 
theory \cite{brogioli2000diffusive}, DFV showed that the observed macroscopic diffusivity $D$ of the 
concentration emerges as an ``eddy-diffusivity'' renormalizing the bare diffusivity $D_0$ predicted 
by kinetic theory and satisfying the Stokes-Einstein relation $D=k_BT/6\pi\eta \sigma,$ with $\eta$ the 
shear viscosity of the solvent and $\sigma$ the molecular radius of a solute molecule. Exploiting 
their high-Schmidt limiting theory both theoretically and numerically, DFV predicted that in free 
diffusion of a blob of mean concentration with initial gradient $|\nabla c_0|$, the long-range non-equilibrium 
correlations in the scalar structure function $S({\bf k},t)$ emerge dynamically through a ``cascade process,'' with 
an initial short-time regime for which $S({\bf k},t)\sim \frac{k_BT}{\eta} |\nabla c_0|^2 t k^{-2}.$ The 
long-time behavior in a quasi-steady regime expected both by experiment 
\cite{vailati1997giant,croccolo2016shadowgraph,vailati2024perspective}
and linearized fluctuating hydrodynamics \cite{vailati1998nonequilibrium} has 
$S({\bf k},t)\sim \frac{k_BT}{D\eta} |\nabla \bar{c}(t)|^2 k^{-4},$ but the numerical dependence 
on wavenumber observed by DFV seemed closer to $k^{-3}.$ This disagreement was rather perplexing
because Donev and his collaborators had previously obtained the expected quasi-steady
result numerically both with Direct Simulation Monte Carlo  \cite{donev2011diffusive,donev2011enhancement} 
and also with nonlinear fluctuating hydrodynamics not simplified by the high-Schmidt limit 
\cite{donev2014low,nonaka2015low}. 

Recently, this discrepancy has been removed by  two of us \cite{eyink2024kraichnan}, who showed
analytically that the DFV theory does indeed yield the expected quasi-steady result in free diffusive mixing. 
However, to our knowledge, this result has not yet been seen numerically in the DFV model 
and, most importantly, it is not understood how the long-range non-equilibrium correlations arise 
dynamically. To address both of these questions, we consider here the simplest example \cite{eyink2024kraichnan}
with initial concentration profile 
\be c_0({\bf x})=\frac{c_0}{2}\left( 1+{\rm erf}\left(\frac{z}{2\sqrt{D\tau}}\right)\right) \lb{init} \ee    
in infinite three-dimensional Euclidean space, which is inhomogeneous only in the $z$-direction. 
The parameter $\tau$ with units of time is introduced to set the magnitude of the initial concentration 
gradient as $|\nabla c_0|=c_0/\sqrt{4\pi D\tau}.$ Physically, the profile \eqref{init} can be 
obtained as the solution of the usual macroscopic diffusion equation with diffusivity $D,$ starting 
from an initial Heaviside step-function profile at a ``virtual time'' origin $-\tau.$
We determine in this example precisely how 
long-range non-equilibrium concentration correlations emerge in free diffusion, by applying novel 
theoretical and computational methods for the DFV model borrowed from the problem of turbulent 
scalar advection \cite{kraichnan1968small,falkovich2001particles}.

\section{\label{sec:dfvt}Quick Review of DFV theory}

Prior to presenting our new methods and results, we briefly review the DFV theory and the 
computational schemes which they had developed. By means of a multiple-scale asymptotic 
method, DFV (and also \cite{eyink2022high}, Appendix B) showed that the nonlinear 
fluctuating hydrodynamic equation for the concentration field in the high-Schmidt limit 
reduces to the Stratonovich stochastic equation 
\begin{equation}
    \partial_t c = -{\bf w} \odot \grad c + D_0\triangle c\,,
    \label{eq:DFV}
\end{equation}
where the thermal velocity field ${\bf w}$ is no longer obtained from a dynamical equation, but 
instead is a Gaussian random field with mean zero and prescribed covariance
\begin{equation}
   \langle {\bf w}({\bf x},t)\otimes  {\bf w}({\bf x}',t') \rangle={\bf R}({\bf x},{\bf x}')\delta(t-t')\,,
    \label{eq:wcov}
\end{equation}
where in terms of the Stokes operator Greens function ${\bf G}$
\begin{equation}
   {\bf R}({\bf x},{\bf x}')=\frac{2k_BT}{\eta}({\bsigma}\star{\bf G}\star {\bsigma}^\top)({\bf x},{\bf x}')\,.
    \label{eq:Rdef}
\end{equation}
Following DFV, we have introduced into the spatial covariance of ${\bf w}$ a low-pass filtering operator 
${\bsigma}$ which smoothly removes wave-numbers $>1/\sigma.$ In fact, nonlinear fluctuating 
hydrodynamic equations such as \eqref{eq:DFV} are generally understood to be ``effective field theories'' 
with an explicit high-wavenumber cut-off 
\cite{zubarev1983statistical,espanol2009microscopic,liu2018lectures} and all of the parameters 
in the model depend upon that cut-off. With wavenumber cut-off $\sim 1/\sigma,$
the parameter $D_0$ has the meaning of ``bare diffusivity'' of the solute. Note that we have 
omitted a molecular noise term in Eq.\eqref{eq:DFV} proportional to $\sqrt{D_0}$ and in fluctuation-dissipation
balance with the diffusion term $D_0\triangle c:$ see \cite{donev2014reversible}, Eq.(4) or 
\cite{eyink2022high}, Eq.(12). As already emphasized by DFV, that noise term is negligible for 
the non-equilibrium phenomena of interest and, in fact, the diffusion term proportional to the 
bare parameter $D_0$ can generally be dropped as well. 

To solve the stochastic equation Eq.\eqref{eq:DFV}, DFV developed two different algorithms.
One approach was an Eulerian numerical scheme, with a staggered finite-volume space-discretization 
and in particular a strictly non-dissipative discretization of the advection term. The velocity
realizations were obtained by solving the steady Stokes equation with a stochastic forcing term 
representing thermal noise, by means of an iterative Krylov linear solver. This  
approach maintains discrete fluctuation-dissipation balance but requires a small, non-vanishing 
$D_0>0$ to dissipate structures finer than the grid created by advection. In this case, the 
bare diffusion term is discretized by a stable, implicit method such as Crank-Nicolson. The 
time-stepping  was performed by an Euler-Heun predictor-corrector scheme which is weakly 
first-order accurate. This high-Schmidt reduction was found to be more efficient by a factor 
proportional to $Sc$ compared with numerical solution of the full non-linear fluctuating hydrodynamic 
equations, which requires solving an extra dynamical equation for the fluctuating velocity field 
and time-resolving the fast viscous dynamics. 

The second numerical method developed by DFV was a Lagrangian approach to solve 
Eq.\eqref{eq:DFV} with $D_0=0$, which represents the solution with initial data $c_0$ 
by the method of characteristics as 
\begin{equation}
    c({\bf x},t)= c_0({\boldsymbol \xi}({\bf x},0))\,,
    \label{c-sol}
\end{equation}
where ${\boldsymbol\xi}({\bf x},t)$ is the Lagrangian flow map that satisfies the Stratonovich 
stochastic differential equation (SDE) 
\begin{equation}
   d{\boldsymbol \xi}({\bf x},s)={\bf W}({\boldsymbol \xi}({\bf x},s),\circ ds), \quad 0<s<t
    \label{xi-eq}
\end{equation}
for ${\bf w}(\bx,t)={\bf W}(\bx,dt)$ and for the final-time condition 
${\boldsymbol \xi}({\bf x},t)={\bf x}.$ (See Appendix \ref{app:mc} for more 
details.) By solving \eqref{xi-eq} 
for a large number $P$ of points $\bx_p,\ p=1,...,P$ suitably distributed 
over the domain, one can get the discrete representation 
\be c({\bf x}_p,t)= c_0({\boldsymbol \xi}({\bf x}_p,0)), \quad p=1,...,P. \ee 
The algorithm was further simplified by DFV using the time-reversibility of the 
Stratonovich flow\cite{kunita1990stochastic} to write 
\begin{equation}
    c({\bf x}_p,t)= c_0({\boldsymbol \xi}({\bf x}_p,t)), \quad  p=1,...,P. 
    \label{c-sol-rev}
\end{equation}
with \eqref{xi-eq} now solved forward in time for $0<s<t$ with ${\boldsymbol \xi}({\bf x}_p,0)={\bf x}_p.$ DFV implemented 
this strategy by generating velocity realizations ${\bf w}$ with a Stokes solver  and by using a predictor-corrector 
scheme to solve the Stratonovich SDE \eqref{xi-eq} forward in time to evaluate \eqref{c-sol-rev}. 
Finally, the It$\bar{{\rm o}}$ and Stratonovich interpretations of \eqref{xi-eq} are equivalent
when $\partial R_{ij}({\bf x},{\bf x})/\partial x_j=0$\cite{kunita1990stochastic},
a condition that easily follows from incompressibility and space-homogeneity.
Thus, in periodic space domains, DFV could solve \eqref{xi-eq} as an It$\bar{{\rm o}}$ 
SDE by the Euler-Maruyama method, generating ${\bf w}$ by a spectral Stokes solver and 
using a nonuniform FFT algorithm to sample only the $M$ values ${\bf w}({\boldsymbol \xi}(\bx_p,s),s)$
for $p=1,...,P$ required to time-advance the SDE. 

\section{New Approaches} 

Unlike DFV, we aim in this work not to calculate full realizations of the concentration field $c$
but instead to calculate only its statistical cumulants, in particular those of second order. Note 
that in a space-homogeneous steady-state, e.g. with constant mean gradients, the concentration 
structure function $S$ is obtained from the second cumulant $C_2({\bf x}-{\bf x}')=C_2({\bf x},{\bf x}')$
by Fourier transform:
\be S({\bf k},t)=\int C_2({\bf r},t) e^{i{\bf k}{\boldsymbol\cdot}{\bf r}} \, d^3r. \lb{FT} \ee
Two methods originating in the theory of turbulent advection are especially 
efficient to compute the cumulants. Both methods apply most directly in physical space. 

\subsection*{\label{sec:closedeq}Solution of closed Eulerian equations for second-cumulant}

A very useful, well-known feature of the Kraichnan model is that there is no {\it closure problem}
\cite{kraichnan1968small,falkovich2001particles}. In particular, cumulants of the concentration field 
of all orders satisfy exact closed equations. 
For instance, the first cumulant or the mean concentration field $C_1(\bx,t)=\langle c(\bx,t)\rangle:=\overline c(\bx, t)$  
satisfies the diffusion equation with a renormalized diffusivity ${\bf D}({\bf x})=D_0{\bf I}+(1/2){\bf R}({\bf x},{\bf x})$:
\begin{eqnarray}\label{C1eq}
&& \partial_t \bar{c}({\bf x}, t)
={1\over 2} 
\nabla_{x^i}\left[{ R}_{ij}(\bx,\bx)\nabla_{x^j} \bar{c}({\bf x}, t) \right] \cr 
&& \qquad \qquad \qquad +D_0 \triangle_{{\bf x}} \bar{c}({\bf x}, t). 
\end{eqnarray}
The second cumulant or concentration covariance 
$C_2({\bf x}_1,{\bf x}_2, t):=  
\langle c({\bf x}_1,t) c({\bf x}_2, t)\rangle-\langle c({\bf x}_1,t)\rangle \langle c({\bf x}_2,t)\rangle$
likewise satisfies a multivariate diffusion equation 
\begin{eqnarray}\lb{C2eq} 
&&\partial_t C_2({\bf x}_1, {\bf x}_2, t)=D_0 \sum_{n=1}^2 \triangle_{{\bf x}_n} C_2({\bf x}_1,{\bf x}_2, t) \cr 
&&\;+{1\over 2} \sum_{n, m=1}^2 
\nabla_{x_n^i}\left[{ R}_{ij}(\bx_n,\bx_m)\nabla_{x_m^j} C_2({\bf x}_1,{\bf x}_2, t) \right]\cr 
&&\; \qquad + R_{ij}({\bf x}_1,{\bf x}_2) \nabla_{x_1^i}\bar{c}({\bf x}_1,t)\nabla_{x_2^j}\bar{c}({\bf x}_2,t)
\end{eqnarray}
with a source term arising from the mean concentration gradient. This pattern repeats 
for all higher orders, each cumulant satisfying a multivariate diffusion equation with 
source determined by lower-order cumulants. 

In this work, we focus on the free diffusion in an unbounded three-dimensional space.
For the initial mean concentration profile \eqref{init}, the equation \eqref{C1eq} 
reduces to the 1D diffusion equation,
\begin{equation}\label{Diff1}
{\partial\over \partial t} \overline c(z, t)=D\partial^2_z\overline c(z, t)
\end{equation} 
where $D=k_BT/6\pi\eta \sigma$ (for $D_0\ll D$), with exact solution 
\bea
\overline c (z, t)={c_0\over 2}\Big(1+{\rm erf}\Big({z\over 2\sqrt{D(t+\tau)}}\Big) \Big). 
\lb{erfc} \eea 
The equation \eqref{C2eq} for the 2nd-order cumulant can also be simplified by introducing 
relative and mean variables
$$ {\bf r}={\bf x}_1-{\bf x}_2=(x,y,z), \quad {\bf X}=\frac{1}{2}({\bf x}_1+{\bf x}_2)=(X,Y,Z), $$
because the velocity statistics are spatially homogeneous and isotropic and thus
${\bf R}$ coincides with the Oseen tensor: 
\begin{equation}
   R_{ij}({\bf r}) = \frac{k_BT}{4\pi \eta r} \left( \delta_{ij} + \frac{r_ir_j}{r^2} \right)\quad \text{if}\quad r\gg\sigma. 
    \label{eq:stokes}
\end{equation} 
The second-cumulant $C_2$ then depends only upon ${\bf r}$ and $Z$ and must furthermore be independent of 
azimuthal angle $\phi$ in spherical coordinates $(r, \theta, \phi).$ 
In those variables, the equation \eqref{C2eq} becomes, for $r\gtrsim \sigma,$
\begin{eqnarray}\nonumber
&& \partial_t C_2
= {k_BT\over 4\pi\eta\sigma}\left(\frac{1}{3}+\frac{\sigma}{4r}(1+\cos^2\theta)\right)
\partial_Z^2 C_2\\\nonumber
&& \quad +{k_BT\over 3\pi\eta\sigma}\cdot {1\over r^2}{\partial\over\partial r}\Big[ r^2\Big(1-{3\over 2}{\sigma\over r}\Big){\partial C_2\over\partial r} \Big]
\\\nonumber
&& \quad +{k_BT\over 3\pi\eta\sigma}\cdot {1\over r^2\sin\theta}\Big(1-{3\over 4}{\sigma\over r}\Big){\partial\over\partial\theta}\Big(\sin\theta{\partial C_2\over\partial\theta}  \Big)\\\label{anisotropic1}
&&\quad + {k_BT\over 4\pi\eta} \cdot {1\over r}(1+\cos^2\theta)
\nabla\overline c\Big(Z+{z\over 2}\Big)\nabla \overline c \Big(Z-{z\over 2}\Big),\qquad
\label{Ceq-slab} \end{eqnarray}
where $r=(x^2+y^2+z^2)^{1/2}$ and $z=r \cos\theta.$ 

The equation \eqref{Ceq-slab} for the second cumulant 
was the basis for the previous calculation of $C_2$ near the mid-plane in the long-time, quasi-steady 
regime\cite{eyink2024kraichnan}:
\bea \label{anisotropic20}
&& C_2(r, \theta,Z,t)\simeq {\mathcal C}^2(t)-{3\over 16} \sigma |\nabla \overline c(t)|^2\;r (2+\sin^2\theta), \cr 
&& \hspace{80pt}  Dt\gg \sigma^2, \quad r\gtrsim\sigma, \quad Z\simeq 0
\eea 
The concentration variance ${\mathcal C}^2(t)=C_2({\bf 0},t)$ cannot be determined from the 
equation \eqref{Ceq-slab}, which is only valid for $r\gtrsim \sigma,$ and its derivation 
would require a regularized version of the equation valid for all $r$ (e.g. see \cite{eyink2022high},
Appendix D). By explicit evaluation of the Fourier transform \eqref{FT}, the physical-space 
result \eqref{anisotropic20} can be shown\cite{eyink2024kraichnan} to be equivalent 
to the standard expression for the concentration structure function of GCF's in free diffusion, 
$S({\bf k},t)\sim \frac{k_BT}{D\eta} |\nabla \bar{c}(t)|^2 |\hat{{\bf k}}_\perp|^2 k^{-4},$
where $\hat{{\bf k}}_\perp$ is the component of $\hat{{\bf k}}={\bf k}/|{\bf k}|$ perpendicular
to $\hat{{\bf z}},$ originally obtained from linearized fluctuating hydrodynamics\cite{vailati1998nonequilibrium}. 

In this work, we derive from \eqref{Ceq-slab} a new physical-space expression for the second cumulant 
in a complementary short-time, large-$r$ regime. Writing the source term in \eqref{Ceq-slab}
for mean concentration \eqref{erfc} explicitly as 
\bea 
&& S(r,\theta,Z,t)= {k_BT\over 4\pi\eta} \cdot {1\over r}(1+\cos^2\theta)
\frac{c_0^2}{4\pi D(t+\tau)} \cr 
&& \qquad \qquad \qquad \times \exp\left[-\frac{Z^2+\frac{1}{4}z^2}{2D(t+\tau)}\right], 
\quad r\gtrsim\sigma \lb{source} \eea 
the key idea is to take this source as the leading-order contribution to the righthand side of \eqref{Ceq-slab}
and treat the diffusion term as a small perturbation. By direct time-integration of the source term \eqref{source}, 
one then obtains 
\bea \lb{larger}
&& C_2(r, \theta,Z,t)\simeq {3 c_0^2\over 8\pi} \cdot {\sigma \over r}(1+\cos^2\theta) \cr 
&& \quad \times \left[ E_1\left( \frac{Z^2+\frac{1}{4}z^2}{2D(t+\tau)}\right) - E_1\left( \frac{Z^2+\frac{1}{4}z^2}{2D\tau}\right) \right],  \cr
&& \qquad \qquad \qquad \qquad Dt\lesssim \sigma^2\ll r^2, 
\eea 
where $E_1(x)$ is the exponential integral function 
(see \cite[\href{https://dlmf.nist.gov/6.2.E1}{6.2.1}]{NIST:DLMF}). 
For more details of the derivation
see Appendix \ref{app:asympt}, 
where it is shown also that the result \eqref{larger} is just the first term in an asymptotic series
for large $r$. 

Note that the expression \eqref{larger} is valid for all $Z.$ Restricting \eqref{larger}
to the special case $Z^2+\frac{1}{4}z^2=0,$ however, the term in square brackets reduces to 
$\ln(1+t/\tau)$ and $\simeq t/\tau$ for $t\ll \tau.$ In fact, using the relation 
$|\nabla c_0|^2=c_0^2/4\pi D\tau,$
\bea \lb{largerZ0}
&& C_2(r, \theta,Z=0,t)\simeq {k_BT\over 4\pi\eta} \cdot t {|\nabla c_0|^2\over r} (1+\cos^2\theta), \cr 
&& \qquad \qquad \qquad \qquad Dt\lesssim \sigma^2 \ll r^2\ll D\tau
\eea 
whence one can see by standard theorems on asymptotics of Fourier transforms 
(see \cite{wong2001asymptotic}, Ch. IX.6, Theorem 4) that this physical-space result 
corresponds to the spectral structure function
\bea \lb{Slarger}
&& S({\bf k},t)\simeq {2k_BT\over \eta} \cdot t |\nabla c_0|^2 {\sin^2\theta_k \over{k^2}}, \cr 
&& \qquad \qquad Dt\lesssim \sigma^2, \ k\sigma\ll 1 \ll Dk^2\tau 
\eea  
where $\sin\theta_k=|\hat{{\bf k}}_\perp|$ is the sine of the angle between 
${\bf k}$ and $\hat{{\bf z}}.$ See Appendix \ref{app:asympt} 
for details. Equation \eqref{Slarger} is exactly the result obtained by DFV, 
\cite{donev2014reversible} section 3.1. In fact, DFV derived their result from a spectral 
version of our Eq.\eqref{C2eq} (see \cite{eyink2000self}, section 2.1 for the equivalence),  
with no mean gradient but with instead non-vanishing spectrum at wavenumber ${\bf k}_0,$ 
so that ${\boldsymbol\nabla}c_0\simeq c_0{\bf k}_0.$ The DFV derivation made transparent 
the analogy of structure-function growth with turbulent cascade. In addition, DFV observed
this short-time regime clearly in their numerical results for the structure function
(see \cite{donev2014reversible}, Figure 3). It is worth noting that, to our knowledge, 
the short-time regime with $k^{-2}$ power-law of the structure function predicted 
by DFV has not yet been observed experimentally. 

\subsection*{\label{sec:lagrangian approach}Lagrangian Monte Carlo evaluation of second-cumulant}

We exploit here as well a Lagrangian numerical scheme like that of DFV, but an even earlier version 
that was developed to calculate statistical correlation functions of a passive scalar 
advected by the Kraichnan model velocity field
\cite{frisch1998intermittency,frisch1999lagrangian,gat1998anomalous}. 
To calculate the $P$th moment function 
$\langle c({\bf x}_1,t)\cdot\cdot\cdot\cdot \ c({\bf x}_P,t)\rangle$ we use
equation \eqref{c-sol-rev} to evaluate $c({\bf x}_p,t)=c_0({\boldsymbol \xi}_p(t)).$
Thus, we evolve $P$ stochastic Lagrangian tracers, ${\boldsymbol \xi}_p(t)={\boldsymbol \xi}({\bf x}_p,t)$, 
$p=1,...,P$ through a vector of correlated Brownian processes, ${\boldsymbol {\mathcal W}}_P(t)=({\bf w}_1(t),...,{\bf w}_P(t))$,  such that ${\bf w}_p(t)={\bf w}({\boldsymbol\xi}_p,t)$. This is implemented with an Euler-Maruyama discretization
\begin{equation}
   {\boldsymbol \xi}_p(t+\Delta t)={\boldsymbol \xi}_p(t)+\tilde{\bf w}_p\sqrt{\Delta t}\,,
    \label{eq:EM}
\end{equation}
where $\tilde{\boldsymbol {\mathcal W}}_P=(\tilde {\bf w}_1,...,\tilde {\bf w}_P)$ is a normal random 
variable having mean zero and $Pd\times Pd$ covariance matrix
\begin{equation}
   R_{pi,qj}=R_{ij}({\boldsymbol \xi}_p(t),{\boldsymbol \xi}_q(t))
   \,,
   \label{eq:R}
\end{equation}
for $p,q = 1,...,P$ and $i,j=1,...,d$ in $d$ space dimensions. The main difference
from the algorithm of DFV is that $P$ is now a small integer ($P=2$ for the 
second cumulant). Thus, rather than a non-uniform FFT, we can generate 
independent samples of the $Pd$-random vector by writing it as 
$\tilde{\boldsymbol {\mathcal W}}_P={\bf L}_P \tilde {\bf N}_P$, where ${\bf L}_P$ 
is the lower triangular Cholesky factor of the positive-definite 
covariance matrix \eqref{eq:R} and $\tilde {\bf N}_P$ is a $Pd$-dimensional 
standard normal vector. 

As in the earlier work on turbulent advection \cite{frisch1998intermittency,frisch1999lagrangian,gat1998anomalous}, 
the ensemble average that defines the second cumulant is approximated by an empirical average 
\be C_2({\bf x}_1,{\bf x}_2,t)\simeq \frac{1}{N} \sum_{n=1}^N c^{\prime(n)}({\bf x}_1,t)c^{\prime(n)}({\bf x}_2,t) 
\lb{eq:sample} \ee 
over a large number $N$ of samples, where the concentration fluctuation field 
is defined by $c^{\prime(n)}({\bf x}_p,t)=c_0({\boldsymbol \xi}_p^{(n)}(t))-\bar{c}({\bf x}_p,t)$ 
and ${\boldsymbol \xi}_p^{(n)}(t),$ $n=,1,...,N$ are independent solutions 
of \eqref{xi-eq} approximated by the Euler-Maruyama scheme \eqref{eq:EM}. The 
convergence of the sample average \eqref{eq:sample} as $N\to\infty$ is slow, with 
typical Monte Carlo errors $\propto 1/\sqrt{N}.$ Nevertheless, the scheme yielded 
accurate results for turbulent advection 
\cite{frisch1998intermittency,frisch1999lagrangian,gat1998anomalous}
with a number of samples at most $N\sim 10^7.$ Unfortunately, 
our problem differs in two crucial respects. First, the covariance for the Kraichnan 
model of a turbulent velocity field had covariance increasing with $r,$ whereas the velocity 
covariance \eqref{eq:stokes} corresponding to thermal noise decreases with $r.$ Formally, the 
thermal velocity field has H\"older exponent $-1/2$ whereas turbulent velocity fields 
have positive H\"older exponents. The consequence is that the elements in the 
$Pd\times Pd$ matrix \eqref{eq:R} off-diagonal in $p,q,$ which are entirely responsible 
for producing correlations between particles, grow smaller in time as particles separate. 
The second key difference is that the turbulence studies 
\cite{frisch1998intermittency,frisch1999lagrangian,gat1998anomalous} considered
a statistical steady-state for the passive scalar with large-scale injection, 
whereas we study the problem of free diffusion for which the concentration correlations decay in time.
This decaying magnitude makes it more difficult to achieve acceptable relative error. 
Both of these differences require that $N$ must be very much larger in our problem.
Fortunately, the calculation of the sample average \eqref{eq:sample} is trivially 
parallelizable, with only the requirement to choose independent seeds for pseudorandom 
number generators on different processors. We have developed and applied a parallel GPU code 
capable to yield up to $N\sim 10^{15}$ independent samples, whose results we 
present below. See Appendix \ref{app:mc} for full discussion of the numerical methods
and their implementation. 

\begin{figure}[htp]
\includegraphics[width=\linewidth]{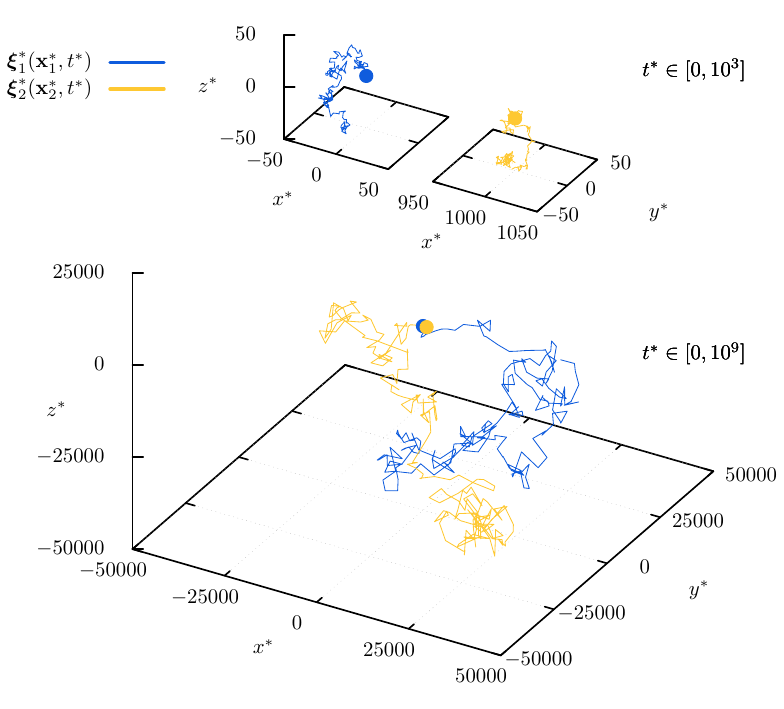}
\caption {\footnotesize {
Trajectories of two Lagrangian tracers ending at the positions ${\bf x}^*_1=(0,0,0)$ and ${\bf x}^*_2=(1000,0,0)$, indicated by the blue and yellow big circles, respectively. The \textit{top} panel is limited to intermediate times, $t^*\le 10^3$, while the \textit{bottom} one extends up to long times, $t^*\le10^9$.
}}
\label{Pair of particles}
\end{figure}

As an illustration of our numerical approach, we plot in Figure \ref{Pair of particles} a 
single pair of Lagrangian particles undergoing correlated Brownian motion according to the 
covariance \eqref{eq:R} that are used to calculate one term in the sum \eqref{eq:sample} 
for the second cumulant. These and all quantities have been computed in dimensionless units with 
$\sigma$, $c_0$, and $\sigma^2/(2D)$ reference scales for length, concentration, and time, respectively. 
In physical units, by considering, for instance, a dispersion of $10\,\rm nm$-sized particles in water at ambient condition, corresponding to $D=2.2\times 10^{-11}\,{\rm m^2/s}$, the reference time is $2.3\times 10^{-6}\,\rm s$.
Non-dimensional variables are hereafter indicated by an asterisk (${\,\!}^*$) superscript. 

\section{\label{sec:numericalresult} Numerical Results}

We here present and discuss our numerical results for the first- and second-order 
cumulants of the concentration fluctuations in the free diffusion of an initial 
mean concentration profile \eqref{init} that corresponds to a planar interface, either 
sharp on the molecular scale ($\tau=0$) or smoothed ($\tau>0$). Our numerical experiment 
is closely analogous to several laboratory experiments\cite{vailati1997giant,croccolo2007nondiffusive}
on free diffusion with similar initial concentration profiles. 
Important differences of these two works from ours are that their spatial domain was finite and 
also gravity acted in the direction perpendicular to the initial interface. Domain size and bouyancy 
effects of gravity are both known to quench the concentration fluctuations at large
scales \cite{vailati1998nonequilibrium}. One experiment \cite{croccolo2007nondiffusive}
created an interface by injecting a concentrated solution into water and by waiting for gravity 
to flatten out large-scale fluctuations: this technique is therefore not able to study 
short-time phenenomena. The other experiment\cite{vailati1997giant} began with a flat 
interface in two immiscible fluids just below their consolution temperature and then rapidly 
raised the temperature above that point so that the two fluids could freely mix. The 
latter experiment in principle might measure short-time fluctuations like those 
derived by DFV and in the previous section.

\subsection*{\label{sec:FOC}First order cumulant}

\begin{figure}[htp]
\includegraphics[width=\linewidth]{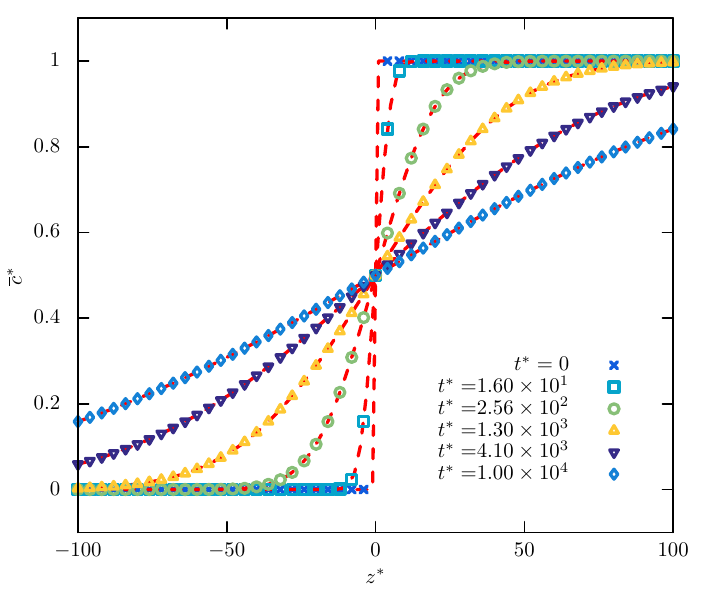}
\caption {\footnotesize {First order cumulant, \textit{i.e.} the mean concentration, versus the distance from the initially sharp ($\tau^*=0$) interface at different times. Symbols represent the numerical results, obtained with $10^9$ samples, while dashed red lines show the analytical solution \eqref{erfc}. Monte Carlo error bars are not visible at the scale of the plot.
}}
\label{First order cumulant}
\end{figure}

As a first simple check of our algorithm, we have computed the time development 
of the mean concentration field for initial profile the sharp step function
given by equation \eqref{init} with $\tau=0.$ We use an $N$-sample average analogous 
to \eqref{eq:sample}:
\be \bar{c}({\bf x},t)\simeq \frac{1}{N} \sum_{n=1}^N c^{(n)}({\bf x},t)
\lb{eq:sample1} \ee 
where now $c^{(n)}({\bf x},t)=c_0({\boldsymbol \xi}^{(n)}(t))$ and 
${\boldsymbol \xi}^{(n)}(0)={\bf x}$ for each $n=1,...,N.$
The results presented in Figure \ref{First order cumulant} show 
excellent agreement with the exact solution of the diffusion equation. 
The plots in Fig.~\ref{First order cumulant} with $t^*\to \tau^*$ represent 
as well the smoothed initial data \eqref{erfc} with $\tau^*>0.$

\subsection*{\label{sec:FOC}Second order cumulant: short-time behavior}

\begin{figure*}[htbp]
  \centering

  \begin{subfigure}[b]{0.49\linewidth}
    \includegraphics[width=1.02\textwidth]{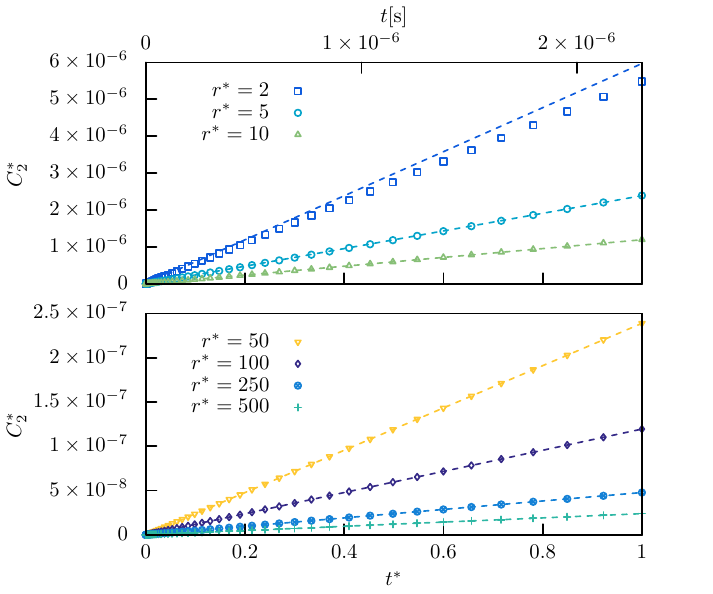}
  \end{subfigure}
  \hfill
  \begin{subfigure}[b]{0.49\linewidth}
    \includegraphics[width=0.95\textwidth]{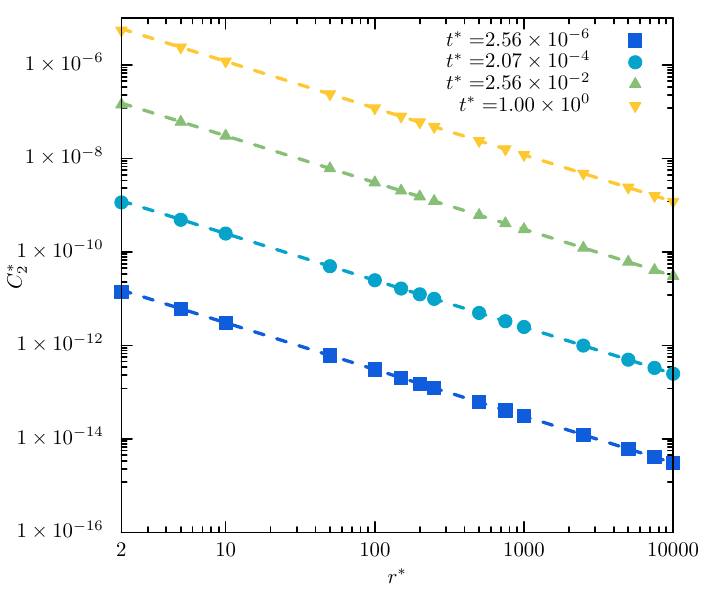}
  \end{subfigure}

  \caption{\footnotesize {
  Second order cumulant in the early time regime for points laying on the interface ($Z^*=0$, $z^*=0$, $\theta=\pi/2$, $\tau^*=10^4$). \textit{Left}: $C_2^*$ versus $t^*$. \textit{Right}: $C_2^*$ versus $r^*$. 
  Different symbols refer to different values of $r^*$ and $t^*$ in the \textit{left} and \textit{right} panels, respectively. Dashed lines represent the analytical prediction in \eqref{largerZ0}.
  The number of samples is $\sim 1.4\times 10^{13}$ and the error bars are not visible at the scales of the plots.
  }}
  \label{fig:In plane second order cumulant short times}
\end{figure*}

\begin{figure*}[htbp]
  \centering

  \begin{subfigure}[b]{0.49\linewidth}
    \includegraphics[width=1.05\textwidth]{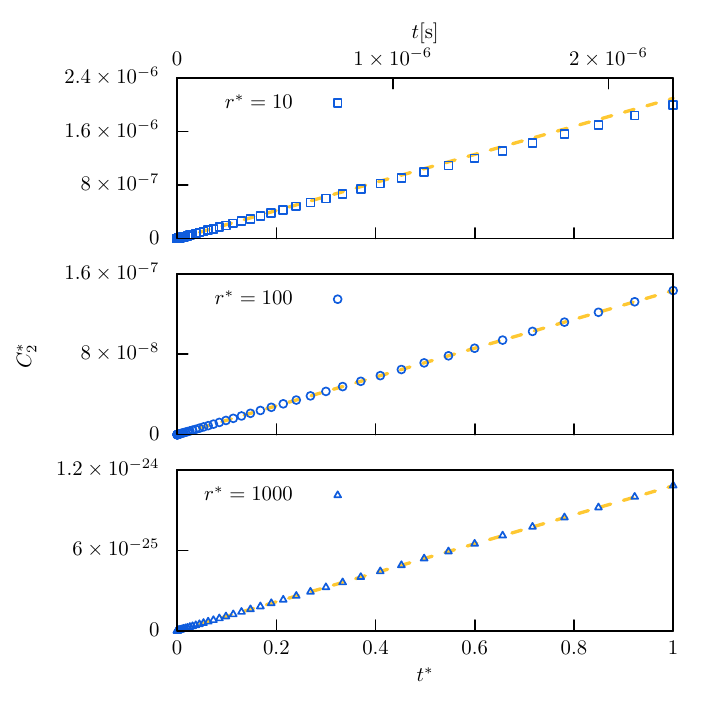}
  \end{subfigure}
  \hfill
  \begin{subfigure}[b]{0.49\linewidth}
    \includegraphics[width=0.95\textwidth]{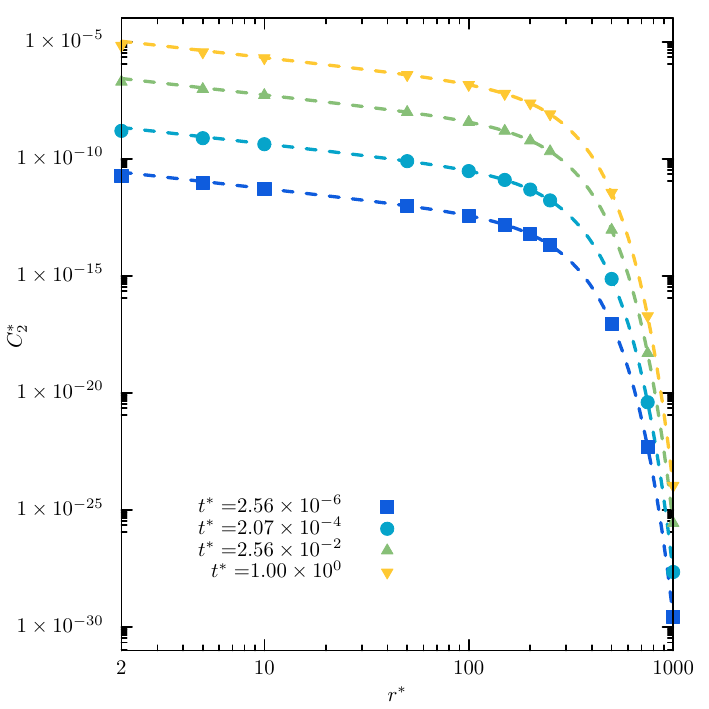}
  \end{subfigure}

  \caption{\footnotesize { 
  Second order cumulant in the early time regime for points laying outside of the interface ($Z^*=r^*\cos (\theta) /2$, $z^*=r^* \cos (\theta)$, $\theta=\pi/6$, $\tau^*=10^4$). \textit{Left}: $C_2^*$ versus $t^*$, with different plots and symbols for different values of $r^*$. \textit{Right}: $C_2^*$ versus $r^*$, with different symbols for different times. 
  Dashed lines represent the analytical prediction in \eqref{larger}.
  The number of samples is $\sim 2.8\times 10^{13}$ and the error bars are not visible at the scales of the plots.
  }}
  \label{fig:Out of plane second order cumulant short times}
\end{figure*}

We next present numerical results for the second cumulant of the concentration 
fluctuations calculated from \eqref{eq:sample1}, first in the short-time,
large-$r$ regime and in the plane of the initial interface for $Z=0,$ in order
to check the prediction \eqref{largerZ0}. We further took $\theta=\pi/2$
so that also $z=0$ and thus the restriction $r^{*2}\ll \tau^*$ in \eqref{largerZ0}
becomes moot.  Finally, we chose $\tau^*=10^4$ so that the approximation should be valid 
for $t^*\lesssim 1$ and $r^*\gg 1.$ The numerical results in Figure \ref{fig:In plane second 
order cumulant short times}, were obtained with numbers of samples $N\sim 1.4\times 10^{13}$ 
and the error bars from Monte Carlo error are smaller than the symbol size to plot the data. 
The numerical values of the cumulant plotted in Figure \ref{fig:In plane second order cumulant short times}(b) 
versus $r^*\in [2,10^4]$ for various times $t^*\in [0,1]$ are in remarkably good agreement 
with the prediction \eqref{largerZ0}. The same results plotted 
in Figure \ref{fig:In plane second order cumulant short times}(a) versus $t^*\in [0,1]$
for various $r^*\in [2,10^4]$ only show a moderate difference in growth for the smallest 
value $r^*=2.$ Thus, the large-$r$ asymptotic result \eqref{largerZ0} agrees with 
the numerical results down to radial distances as small as $r^*=5.$

As previously discussed, the physical-space result \eqref{largerZ0} is equivalent 
to the prediction \eqref{Slarger} of DFV for the spectral structure function. 
However, our more general short-time, large-$r$ result \eqref{larger} applies 
also out of the mid-plane $Z=0.$ To test this more general prediction, we plot in 
Figure \ref{fig:Out of plane second order cumulant short times} numerical results for 
$Z^*=r^*\sqrt{3}/4$ and $\theta=\pi/6,$ obtained with numbers of samples $N\sim 2.8\times 10^{13}$ 
and Monte Carlo error bars too small to be observed. The prediction \eqref{larger} should hold for 
$t^*\lesssim 1$ and $r^*\gg 1,$ so that we plot in Fig.\ref{fig:Out of plane second order cumulant short times}(b)
the cumulant values versus $r*\in [2,10^3]$ for various times $t^*\in [0,1].$ 
These are again in remarkably good agreement with the prediction \eqref{larger}. 
The same results plotted in Fig.\ref{fig:Out of plane second order cumulant short times}(a) 
versus $t^*\in [0,1]$ for various $r^*\in [10,10^3]$ show a slight difference in growth 
only for the smallest value $r^*=10$ but otherwise agree with 
with the asymptotic prediction \eqref{larger}. To our knowledge, none of these short-time 
results have yet been observed in laboratory experiment.

\subsection*{\label{sec:FOC}Second order cumulant: long-time behavior}

\begin{figure*}[htbp]
  \centering

  \begin{subfigure}[b]{0.49\linewidth}
    \includegraphics[width=\textwidth]{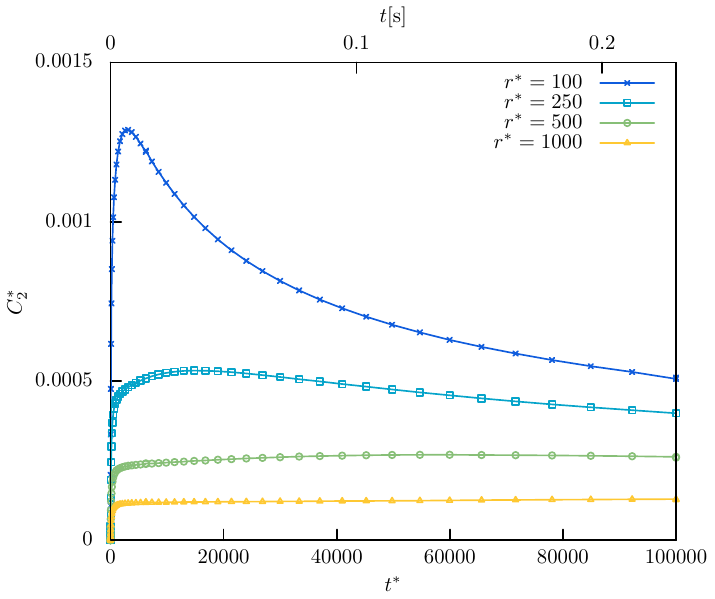}
  \end{subfigure}
  \hfill
  \begin{subfigure}[b]{0.49\linewidth}
    \includegraphics[width=\textwidth]{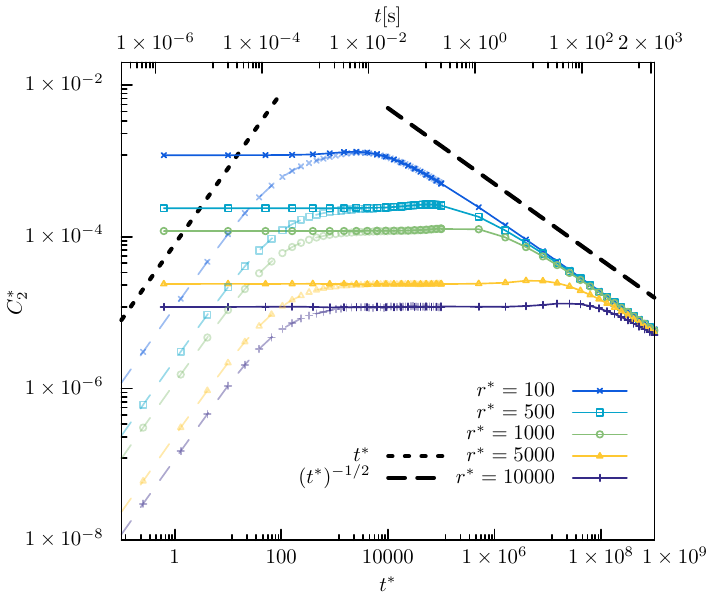}
  \end{subfigure}

  \caption{\footnotesize {
  Second order cumulant versus time (full timescale) for points laying on the interface ($Z^*=0, z^*=0, \theta=\pi/2$). Different symbols represent different values of $r^*$ in both panels. (\textit{Left}) From early to intermediate times with initially smooth interface ($\tau^*=10^2$). (\textit{Right}) Superposition of results from initially sharp ($\tau^*=0$) and initially smooth interfaces ($\tau^*=10^2$) in solid and translucent lines with symbols, respectively. Power laws are also reported in black dashed lines.
  Data are obtained by averaging on $1.5\times 10^{13}$ samples and the error bars are too small to be seen at this scale.
  }}
  \label{fig:Second order cumulant long times}
\end{figure*}

\begin{figure}[htp]
\includegraphics[width=\linewidth]{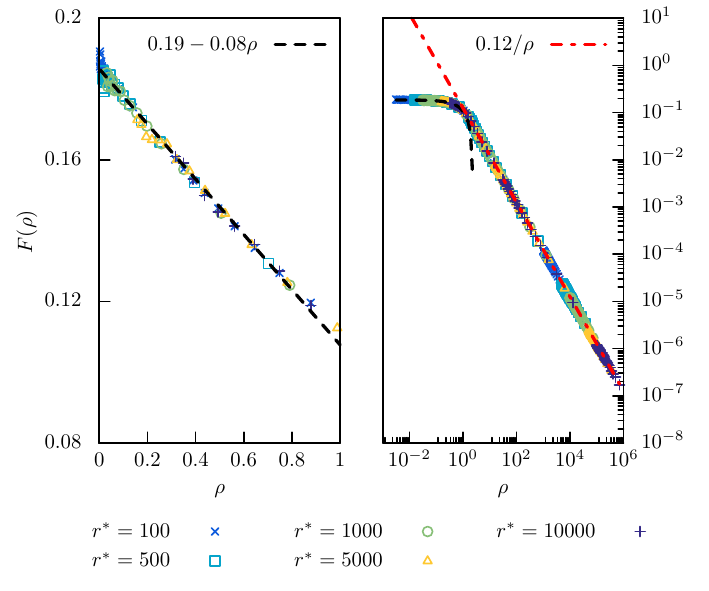}
\caption {\footnotesize {
    Self-similarity of the second order cumulant for points laying on the initially sharp interface ($Z^*=0, z^*=0, \theta=\pi/2$, $\tau^*=0$). 
    The scaling function $F(\rho)=C_2(r^*,t^*)/{\mathcal C}^2(t^*)$ is shown versus $\rho$ with different symbols for different values of $r^*$. The black dashed line is a linear fit for $\rho\le 1$ and the red dot-dashed line is the fit of a $\propto 1/\rho$ function for $\rho> 1$.
    The number of samples is $1.5\times 10^{13}$. For the sake of clarity, error bars are omitted here. However, they do not alter the scaling law.
}}
\label{Second order cumulant long time scaling function}
\end{figure}

\begin{figure*}[htbp]
  \centering

  \begin{subfigure}[b]{0.49\linewidth}
    \includegraphics[width=\textwidth]{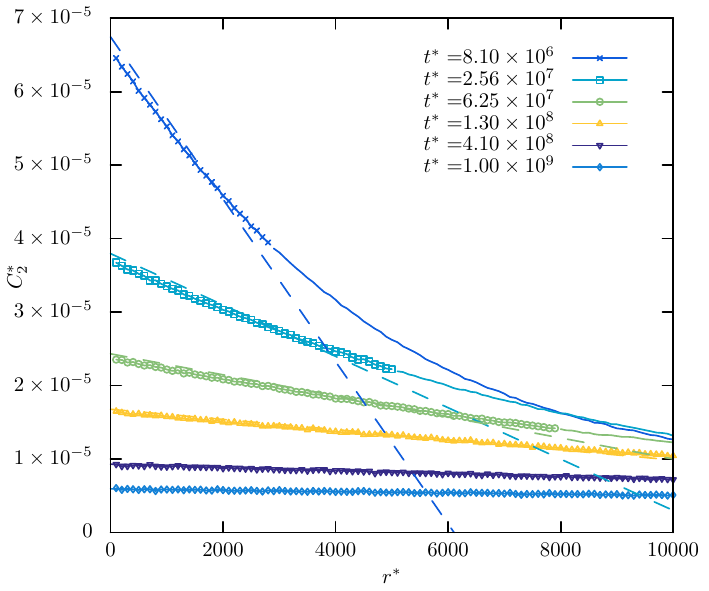}
  \end{subfigure}
  \hfill
  \begin{subfigure}[b]{0.49\linewidth}
    \includegraphics[width=\textwidth]{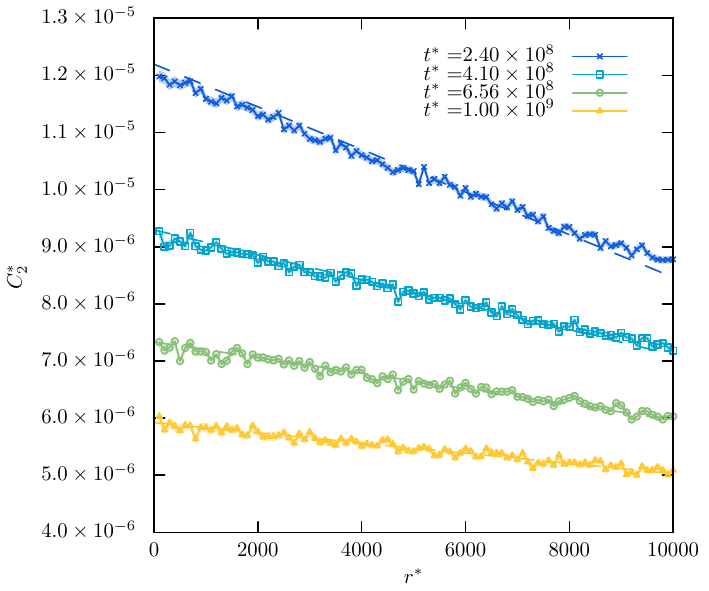}
  \end{subfigure}

  \caption{\footnotesize {
  Long-time regime of the second order cumulant versus the separation $r^*$ for points laying on the interface ($Z^*=0$, $z^*=0$, $\theta=\pi/2$, $\tau^*=0$). Different solid lines represent different times, with symbols shown only for $\rho\leq1$. Dashed lines show the expected linear behavior with given slope \eqref{anisotropic20} and fitted variance. \textit{Left}: from intermediate to long times. \textit{Right}: enlargement on long times.
  Data are obtained with $1.5\times 10^{13}$ samples and error bars are shown by the shaded bend around solid lines (best visible in the \textit{right} panel).
  }}
  \label{fig:Second order cumulant vs R long times}
\end{figure*}

A very interesting question is how the long-time quasi-steady regime \eqref{anisotropic20}, 
seen in laboratory experiments on free diffusion\cite{vailati1997giant,croccolo2007nondiffusive},
emerges from the early-time growth regime discussed in the previous section. Note that the 
early-time prediction was obtained by expanding in $t/\tau$ and its existence depends upon 
the assumption of an initial graded interface with $\tau>0.$ Furthermore, 
the theoretical derivation\cite{eyink2024kraichnan} of the quasi-steady prediction \eqref{anisotropic20}
did not derive the time-dependence of the concentration variance $\mathcal{C}^2(t).$ We address here 
these fundamental issues, not yet answered theoretically, with our numerical 
Monte Carlo results. For the case $\tau^*=10^2,$ Figure \ref{fig:Second order cumulant long times}(a)
plots the concentration second cumulants obtained with a number of samples $N=1.5\times 10^{13}$ 
for four values of $r^*=100,250,500,1000$ for a longer range of times $t^*\in [0,10^5].$ The 
cumulant values initially rise linearly, as predicted, but then achieve a maximum value and thereafter 
decay slowly at long times. We note that this maximum value at intermediate times corresponds to the largest 
experimental signal and may be accessible to laboratory measurement. To better determine the long-time 
behavior we plot these results on a much longer time-interval $t^*\in [10^{-1},10^9]$ with a log-scale 
and including additional values $r^*=5000,10000$ in Figure \ref{fig:Second order cumulant long times}(b). 
The early-time growth indicated by the dotted line is confirmed to be $\propto t^*,$ as expected, and 
furthermore the late-time decay indicated by the dashed line is observed to be $\propto (t^*)^{-1/2}.$ 
We plot also in this figure the corresponding results for the case of an initial concentration 
interface which is sharp on the molecular scale, with $\tau=0.$ This case lacks the early-time 
linear growth regime, but its late-time behavior coincides with that for $\tau>0.$ For simplicity,
we hereafter consider the late-time regime mainly for the case $\tau=0.$

Since the term $\propto r$ in the quasi-steady prediction \eqref{anisotropic20} decays 
in time at the faster rate $|\grad c(z=0,t)|^2\propto 1/t,$ we can thus identify the origin
of the slow decay with ${\mathcal C}^2(t)\propto t^{-1/2}.$ This power law can be 
explained if one postulates a long-time self-similar decay, at least in the central plane 
perpendicular to the gradient:
\be C_2(r,\theta=\frac{\pi}{2},Z=0,t) \sim {\mathcal C}^2(t) 
    F(r/L(t)). \lb{self-sim} \ee 
Adopting the reasonable hypothesis that $L(t)\propto (Dt)^{1/2},$ then 
\eqref{anisotropic20} can be consistent with the self-similar ansatz \eqref{self-sim} only 
if ${\mathcal C}^2(t)\propto c_0^2 \sigma/(Dt)^{1/2}.$ 

The numerical results for the 
scaling function $F(\rho):= C_2(r,\theta=\frac{\pi}{2},Z=0,t)/{\mathcal C}^2(t)$
as a function of $\rho=r/L(t)$ with $\tau=0$ are plotted in 
Figure \ref{Second order cumulant long time scaling function} for five different 
values of $r^*.$ The very good collapse of the results 
for the different values of $r^*$ supports (and indeed originally suggested) 
the hypothesis of asymptotic self-similarity as in \eqref{self-sim}. We observe
in Figure \ref{Second order cumulant long time scaling function}(a) for the range 
$\rho<1$ a linear behavior of $F(\rho),$ with the best fit indicated by the dashed line, that 
corresponds to the experimentally observed ``giant concentration fluctuations''
with the structure function $\propto k^{-4}.$ Interestingly, 
Figure \ref{Second order cumulant long time scaling function}(b) for the 
range $\rho\gg 1$ with a log-scale shows instead an asymptotic behavior $\propto 1/\rho,$
with the best fit indicated by the dash-dotted line. To our knowledge, this quasi-steady 
regime, which would correspond to a structure function $\propto k^{-2}$ at very low wavenumbers, 
has never been observed experimentally, presumably due to domain-size effects. 

The appearance of self-similar behavior at long times should not be unexpected, 
since it is typical in turbulent decay processes. 
To demonstrate it analytically, one could follow standard 
approaches for turbulently advected passive scalars\cite{eyink2000self}.
There are, however, two important differences in the present case
compared with turbulent scalars. First, the model fields for a turbulent 
velocity are chosen to have spatial H\"older exponent $h$ in the range 
$0<h<1,$ whereas the thermal velocity field with spatial covariance 
given by the Oseen tensor \eqref{eq:stokes} has negative H\"older
exponent $-1/2.$ The resulting equation \eqref{C2eq} for $C_2$
thus is more similar to the equation for the correlation function 
of a turbulently advected magnetic field, where the stretching term 
in the magnetic induction equation involves the velocity gradient 
with negative H\"older exponent $h-1.$ Self-similar decay solutions have been
studied\cite{eyink2010small} for the magnetic correlation function 
in the non-dynamo regime of the Kraichnan-Kazantsev model, but this required
including a resistive regularization and using a matched asymptotic expansion 
to patch together solutions at the inertial scale and at the resistive scale. 
Similarly in our problem, a regularization of the equation \eqref{Ceq-slab} 
at scales $r<\sigma$ (e.g. see \cite{eyink2022high}, Appendix D) must be 
introduced and matched solutions at those scales and at $r>\sigma$ must 
be developed. The second additional difficulty in the current 
problem is that statistics are spatially inhomogeneous and anisotropic.  
This problem may be addressed by considering a partially integrated 
correlation function 
\be \bar{C}_2(r,t):=\int_{-\infty}^{+\infty} dZ \int_0^\pi \sin\theta \, d\theta\, 
C_2(r,\theta,Z,t) \ee 
which is now homogeneous and isotropic and which satisfies a simplified 
equation 
\be
\partial_t \bar{C}_2 = {k_BT\over 3\pi\eta\sigma}\cdot {1\over r^2}{\partial\over\partial r}\Big[ r^2\Big(1-{3\over 2}{\sigma\over r}\Big){\partial \bar{C}_2\over\partial r} \Big] + \bar{S}(r,t) \label{Ceq-slab-avrg} \ee
where $\bar{S}$ is obtained by similarly integrating the source term $S$ in equation \eqref{source}:
\bea 
&& \hspace{-10pt} \bar{S}(r,t)={k_BT\over 4\pi\eta} \cdot {c_0^2\over r^2} 
\Bigg[\left(1+\frac{4D(t+\tau)}{r^2}\right){\rm erf}\left(\frac{r}{\sqrt{8D(t+\tau)}}\right)\cr
&& \hspace{60pt} -\frac{(2D(t+\tau))^{1/2}}{r} \exp\left(\frac{-r^2}{8D(t+\tau)}  \right)\Bigg] 
\eea
As in the quasi-steady regime\cite{eyink2024kraichnan}, solving this 
equation can be the first step to obtain the full solution in the 
asymptotic self-similar regime. We hope to pursue this problem in future work.

Returning to the numerical results, we confirm in detail the quasi-stationary
prediction \eqref{anisotropic20} at long times and at length scales $\sigma\ll r\ll L(t).$
In Figure \ref{fig:Second order cumulant vs R long times} we plot our numerical results 
for the second cumulant versus $r^*$ for very long times. We see 
in Fig.~\ref{fig:Second order cumulant vs R long times}(a) that the quasi-steady predictions 
hold well for $t^*\gtrsim 10^7$ over a growing range of lengths $r^*,$ as expected. 
Note that the linear fits have intercepts determined by the empirical decay law for ${\mathcal C}^2(t)$
but the slopes are given analytically by the predicted quasi-stationary formula \eqref{anisotropic20}.
To emphasize the quality of the fit at long times, we plot in 
Fig.~\ref{fig:Second order cumulant vs R long times}(b) the numerical results 
for four times $t^*>10^8$ over the range $r^*\in [0,10^4]$ where one can see 
the linear scaling with the predicted slope matches the data very well. 

\begin{figure}[htbp]
\includegraphics[width=\linewidth]{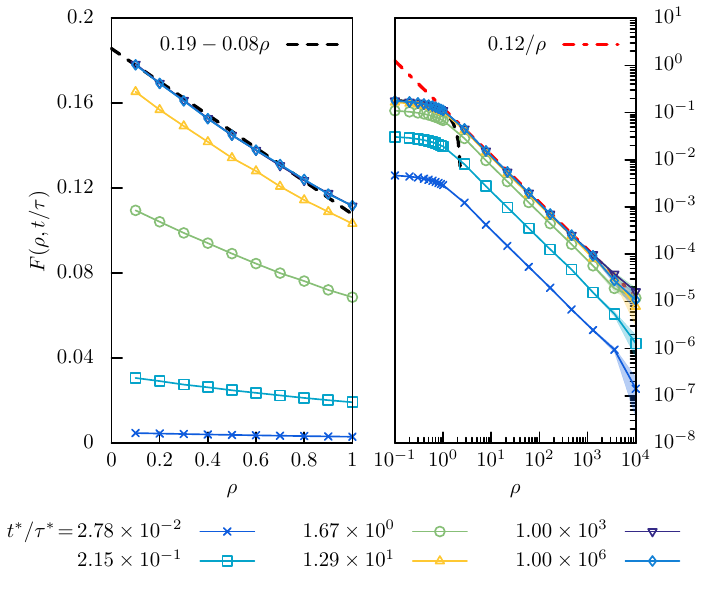}
\caption {\footnotesize {  
    Dynamical development of the self-similar second order cumulant for points laying on an initially smooth interface ($Z^*=0, z^*=0, \theta=\pi/2$, $\tau^* >0$).  
    The scaling function $F(\rho)=C_2(r^*,t^*)/{\mathcal C}^2(t^*)$ is shown versus $\rho$ with different symbols now representing different normalized times, $t^*/\tau^*$. The black dashed and the red dot-dashed lines are taken from Fig.~\ref{Second order cumulant long time scaling function} and represent the expected behaviors for $t^*/\tau^* \to \infty$ when $\rho < 1$ and $\rho >1$, respectively.
    The number of samples varies for each point shown, with the most demanding conditions (large $\rho$) requiring $1.5\times 10^{13}$ realizations. All results are obtained by varying $\tau^*$, $t^*$, and $r^*$ while guaranteeing that $r^*\ge 50$. Colored bends show the Monte Carlo error. 
}}
\label{Second order cumulant approach to scaling function}
\end{figure}

Finally, we study the very fundamental issue of the emergence of the long-range 
correlations and the approach to the asymptotic self-similar behavior at long times.
In order to provide an initial regime before self-similiarity develops, we consider 
the case $\tau>0$ and plot in Figure \ref{Second order cumulant approach to scaling function}
the results for $F(\rho,t/\tau):=
C_2(r,\theta=\frac{\pi}{2},Z=0,t)/{\mathcal C}^2(t)$ versus $\rho=r/L(t)$ at six values of the 
dimensionless time $t/\tau=t^*/\tau^*.$ For comparison, we plot also the fit for the asymptotic 
scaling function $F(\rho)$ determined from Fig.~\ref{Second order cumulant long time scaling function}. 
As expected from our short-time, large-$r$ asymptotic analysis, the tail $F(\rho)\sim 0.12/\rho$ 
at $\rho\gtrsim 1$ is confirmed in Fig.~\ref{Second order cumulant long time scaling function}(b)
to be approached from below by a similar $1/\rho$-tail with amplitude initially growing 
linearly in $t/\tau.$ Very interestingly, the emergence of the standard quasi-stationary 
regime with linear behavior $F(\rho)\sim 0.19-0.08\rho$ for $\rho\lesssim 1$ is seen 
in Fig.~\ref{Second order cumulant long time scaling function}(a) to emerge in a similar 
manner, so that the results $F(\rho,t/\tau)$ are described well by affine functions with 
negative slopes $\sim t/\tau$ at early times. To our knowledge, this emergence 
of the asymptotic quasi-steady behavior has not been observed previously either by 
simulation or by experiment.

\section{\label{sec:dc}Discussion and Conclusions}

The main contribution of this work has been to study the DFV model\cite{donev2014reversible} of 
high-Schmidt mixing 
using tools borrowed from turbulence theory. First, we have exploited the existence
of a closed hierarchy of dynamical equations for cumulants of all orders which may, in principle, 
be solved iteratively. Extending the short-time approximation of DFV to large 
spatial distances, we have obtained an exact asymptotic form for the second cumulant 
of concentration fluctuations in physical space. In particular, our formula extends 
the result of DFV to points outside the mid-plane at $Z=0.$ We have furthermore applied 
to the DFV model a Lagrangian Monte Carlo numerical method to calculate the cumulants. 
This approach verifies the predictions of DFV and ourselves for the short-time, large-$r$ regime.
Also, and apparently for the first time, we have observed numerically in the DFV model 
the standard long-time quasi-stationary regime in free diffusion from an initial mean 
profile, with non-equilibrium long-range correlations $\propto r$ for $r\lesssim L(t),$
where $L(t)=(2Dt)^{1/2}.$
More importantly, our numerical study has discovered a new quasi-steady regime at scales $r\gtrsim L(t)$ 
with decaying correlation of concentration fluctuations $\sim 1/r$.  
Furthermore, the two ranges for $r\lesssim L(t)$ and for $r\gtrsim L(t)$ appear together
in an asymptotic self-similar decay solution. Finally, we have established exactly how 
this self-similar form is approached in time, thus answering for high-Schmidt liquid 
mixing the fundamental question how non-equilibrium long-range correlations emerge. 
It would be very illuminating to study this transient process in laboratory 
experiments in order to determine the accuracy of the model predictions. 

We learned much about this subject from conversations with Aleks Donev, 
who warmly encouraged us to pursue the connection between liquid diffusion
and turbulent cascade. Aleks was especially excited about the time-reversibility 
of the DFV model, fully manifest when taking the bare diffusivity $D_0=0$. 
He was passionate about the vision of high-Schmidt scalar mixing as an effect of 
advection by thermal velocity fluctuations in detailed balance, going beyond traditional 
Fickian diffusion, and he believed that this more accurate portrayal offered fundamental 
new perspectives into the foundations of hydrodynamics. We are honored to offer this work 
in memory of Aleks and his great legacy in the study of fluctuating hydrodynamics, colloids,
materials science, and computational biology. 

\textit{Data Availability.} 
The data that support the findings of this study will be available 
through a public repository on the Zenodo platform.

\textit{Author Declarations.} 
The authors have no conflicts to disclose. 

\begin{acknowledgments}

We are grateful to many colleagues for discussions of this work, most especially Aleksandar Donev, 
and including also Roberto Benzi, Luca Biferale, Carlo Massimo Casciola, Mauro Sbragaglia, and Eric Vanden-Eijnden. 
This work is supported by an ERC grant (ERC-STG E-Nucl. Grant agreement ID: 101163330) (PI M. Gallo).
Support is acknowledged also from the Sapienza Funding Scheme ``Avvio alla Ricerca - Tipo 2", project No. AR22419078AD8186 (PI M. Bussoletti). Computational resources were made available from ICSC-Italian Research Center on High Performance Computing, Big Data, and Quantum Computing under ``MDR-TP - Spoke 6". We also acknowledge support for computational resources from CINECA under the ISCRA iniative, relative to the ISCRA-B D-RESIN (PI M. Gallo) and ISCRA-C EMADON (PI M. Bussoletti) projects. 
G. Eyink thanks the Department of Physics of the University of Rome ‘Tor Vergata’ for hospitality while this work was begun and acknowledges support from the European Research Council (ERC) under the European Union’s Horizon 2020 research and innovation program (Grant Agreement No. 882340). {\it Funded by the European Union}. Views and opinions expressed are however those of the authors only and do not necessarily reflect those of the European Union or the European Research Council Executive Agency. Neither the European Union nor the granting authority can be held responsible for them. 
\end{acknowledgments}

\appendix

\section{Short-time, Large-distance Asymptotics}\lb{app:asympt}

The zeroth order approximation \eqref{larger} to the second cumulant at small $t$
and large $r$ is given by the time-integral of the source \eqref{source}: 
\be C_2^{(0)}(r, \theta,Z,t) = \int_0^t S(r, \theta,Z,s) \, ds \lb{C2_0} \ee
By the change of variables $u=a/(s+\tau)$
\bea && \int_0^t \exp\left(-\frac{a}{s+\tau}\right) \frac{ds}{s+\tau}
     =\int_{a/(t+\tau)}^{a/\tau} e^{-u} \frac{du}{u} \cr 
     && \hspace{50pt} =E_1\left( \frac{a}{t+\tau}\right) - E_1\left( \frac{a}{\tau}\right) \eea 
where $E_1(x)=\int_x^\infty e^{-u} \, du/u$ is the exponential integral function 
(see \cite[\href{https://dlmf.nist.gov/6.2.E1}{6.2.1}]{NIST:DLMF})
Using this result to evaluate \eqref{C2_0} 
yields immediately \eqref{larger} in the main text. 
Because $E_1(x)\sim e^{-x}/x$ for $x\to\infty$
(see \cite[\href{https://dlmf.nist.gov/6.12.E1}{6.12.1}]{NIST:DLMF}), 
we have the upper bound $C_2^{(0)}=O\left(\frac{t}{\tau}\frac{\sigma}{r}\right).$ 

This approximation is just the leading term in an asymptotic expansion 
for small $t$ and large $r$: 
\be C_2(r, \theta,Z,t) =C_2^{(0)}(r, \theta,Z,t) + C_2^{(1)}(r, \theta,Z,t) + \cdots 
\lb{asympt} \ee 
Substituting \eqref{asympt} into \eqref{Ceq-slab} yields an equation for $C_2^{(1)}$: 
\begin{eqnarray}\nonumber
&& \partial_t C_2^{(1)}
= {k_BT\over 4\pi\eta\sigma}\left(\frac{1}{3}+\frac{\sigma}{4r}(1+\cos^2\theta)\right)
\partial_Z^2 C_2^{(0)}\\\nonumber
&& \quad +{k_BT\over 3\pi\eta\sigma}\cdot {1\over r^2}{\partial\over\partial r}\Big[ r^2\Big(1-{3\over 2}{\sigma\over r}\Big){\partial C_2^{(0)}\over\partial r} \Big]
\\\nonumber
&& \quad +{k_BT\over 3\pi\eta\sigma}\cdot {1\over r^2\sin\theta}\Big(1-{3\over 4}{\sigma\over r}\Big){\partial\over\partial\theta}\Big(\sin\theta{\partial C_2^{(0)}\over\partial\theta}  \Big) \label{anisotropic1}
\end{eqnarray}
The contributions from derivatives $\partial_r$ and $\partial_\theta$ straightforwardly 
give $\partial_t C_2^{(1)}= O\left(\frac{t}{\tau}\frac{\sigma}{r}\frac{D}{r^2}\right).$ This 
is less obvious for contributions arising from the derivative $\partial_Z$ but an explicit computation shows that 
\bea  
&& D \partial_Z^2 E_1\left(\frac{Z^2+\frac{1}{4}z^2}{2D\tau}\right) \cr 
&& = 2D \left(\frac{Z^2}{D\tau}-1\right)\frac{\exp\left(-(Z^2+\frac{1}{4}z^2)/2D\tau\right)}{Z^2+\frac{1}{4}z^2} \cr 
&& \hspace{40pt} +4DZ^2 \frac{\exp\left(-(Z^2+\frac{1}{4}z^2)/2D\tau\right)}{\left(Z^2+\frac{1}{4}z^2\right)^2}
\eea 
and likewise for $\tau\to t+\tau.$ Since $z=r\cos \theta,$ these contributions satisfy the same bound 
as the others. Hence, integrating in time, $C_2^{(1)}=O\left(\frac{t}{\tau}\frac{\sigma}{r}\frac{Dt}{r^2}\right).$ 
Computation of further terms in the asymptotic expansion \eqref{asympt} proceeds in the same 
manner, with the result that $C_2^{(k)}=O\left(\frac{t}{\tau}\frac{\sigma}{r}\left(\frac{Dt}{r^2}\right)^k\right)$
for all integers $k\geq 0.$

Using the convergent expansion (see \cite[\href{https://dlmf.nist.gov/6.6.E2}{6.6.2}]{NIST:DLMF})
\be E_1(x) = -\gamma -\ln x -\sum_{k=1}^\infty \frac{(-x)^k}{k\cdot k!} \ee
it follows that 
\be \lim_{a\to 0} [E_1(ax)-E_1(ay)] = \ln(y/x). \lb{log-id} \ee 
For the case $Z^2+\frac{1}{4}z^2=0$ this identity shows that the term in the 
square bracket of \eqref{larger} reduces to $\ln(1+t/\tau).$ Thus \eqref{log-id} yields
\eqref{largerZ0} for the special case $Z=0,$ $\theta=\pi/2.$ More generally,
one can derive \eqref{largerZ0} for $\theta\neq \pi/2$ by considering the time-derivative at $t=0$
\bea 
&& \left.\partial_t C_2(r, \theta,Z=0,t)\right|_{t=0} = S(r, \theta,Z=0,t=0)\cr
&& \hspace{-10pt} = {k_BT\over 4\pi\eta} {|\nabla c_0|^2\over r} (1+\cos^2\theta) 
\times \exp\left[-\frac{r^2\cos^2\theta}{8D\tau}\right] \lb{tderC2Z0}
\eea 
The exponential factor in \eqref{tderC2Z0} is $\simeq 1$ for $D\tau\gg r^2,$ 
so that \eqref{largerZ0} then emerges as the first-order Taylor approximation in time.

The Fourier transform of \eqref{largerZ0} (with our convention in equation \eqref{FT}) 
can be evaluated from standard asymptotics of Fourier integrals 
(see \cite{wong2001asymptotic}, Ch. IX.6, Theorem 4): 
$$ {\mathcal F}\left(\frac{1}{r}+\frac{z^2}{r^3}\right) \ \simeq \ 
(2\pi)^{3/2}\left[\frac{L}{k^2}-L^*\frac{\partial^2}{\partial k_z^2}(\ln k)\right]$$
where ${\mathcal F}$ denotes Fourier transform, while 
$$L=2^{3/2-1}\Gamma\left(\frac{3-1}{2}\right)\Big/\Gamma\left(\frac{1}{2}\right)=\left(\frac{2}{\pi}\right)^{1/2}$$
and 
$$L^*= -1\Big/2^{3/2-1}\Gamma\left(\frac{3}{2}\right)=-\left(\frac{2}{\pi}\right)^{1/2}.$$
Since $\frac{\partial^2}{\partial k_z^2}(\ln k)=\frac{1}{k^2}-\frac{2k_z^2}{k^4},$ we obtain
$$ {\mathcal F}\left(\frac{1}{r}+\frac{z^2}{r^3}\right) \simeq  8\pi\left(\frac{k^2-k_z^2}{k^4}\right)
=8\pi \frac{\sin^2 \theta_k}{k^2}. $$
Applying this result to the correlation in \eqref{largerZ0} we obtain immediately the 
Fourier structure function \eqref{Slarger}. 

\section{Monte Carlo Numerical Solver}\lb{app:mc} 

The stochastic Lagrangian flow is defined by equation \eqref{xi-eq} in terms of the 
Brownian vector field 
\be {\bf W}(\bx,t) = \sum_{n=0}^\infty \lambda_n^{-1/2} {\boldsymbol \phi}_n(\bx) W_n(t), \ee
where ${\boldsymbol \phi}_n(\bx),$ $n=0,1,2,...$ are the complete set of orthonormal
eigenfunctions of the Stokes operator, $\lambda_n>0,$ $n=0,1,2,...$ are the corresponding
eigenvalues, and $W_n(t),$ $n=0,1,2,...$ are a set of i.i.d. standard Brownian motions.
The zero-mean, white-noise field ${\bf w}(\bx,t)={\bf W}(\bx,dt)$ thus has the covariance 
\eqref{eq:wcov} and the flow equation \eqref{xi-eq} can be written in more detail as
the Stratonovich stochastic differential equation (SDE): 
\begin{equation}
   d{\boldsymbol \xi}({\bf x},t)=\sum_{n=0}^\infty \lambda_n^{-1/2} {\boldsymbol \phi}_n({\boldsymbol \xi}({\bf x},t))
     \circ {\bf W}_n(dt)
    \label{xi-eq2}
\end{equation}
Note, however, that converting this equation to It$\bar{\rm o}$ form merely adds 
a drift field $\frac{1}{2} \left.\partial R_{ij}(\bx,\bx')/\partial x_j\right|_{\bx'=\bx},$ 
which vanishes for the current problem because of the space-homogeneity and 
incompressibility of the random field ${\bf w}(\bx,t).$ For this reason, the 
Lagrangian Monte Carlo algorithm presented in Section \ref{sec:lagrangian approach}
can employ the Euler-Maruyama discretization \eqref{eq:EM} which converges to the 
It$\bar{\rm o}$ SDE's for the $p$ correlated Lagrangian particles obtained 
by taking $\bx=\bx_k,$ $k=1,...,p$ in \eqref{xi-eq2}. For more background, 
see standard texts on stochastic flows\cite{kunita1990stochastic}.

The Monte Carlo algorithm presented in Section \ref{sec:lagrangian approach} is based on a regularized version of \eqref{eq:R} that accounts for the filtering of wavenumbers $\gtrsim 1/ \sigma$. Specifically, we implement the exponentially decaying kernel proposed in \cite{eyink2022high}, Appendix D, that also reproduces the expected Stokes-Einstein macroscopic diffusion $D=k_BT/6\pi\eta \sigma$ in \eqref{Diff1}. For the sake of better performance, the numerical solver implements the regularization when $|{\bf x}_1 - {\bf x}_2|\le10\sigma$ and \eqref{eq:stokes} otherwise. Moreover, we exploit an adaptive time-stepping method for the Euler-Maruyama time-discretization in \eqref{eq:EM} such that $\Delta t= (f\, |{\bf x}_1 - {\bf x}_2|)^2/2D$, where $0<f\le1$ is a parameter controlling what fraction of the distance between the two Lagrangian tracers can be covered on average with one time-step. We use $f=0.2$ in our results -- tests with smaller and larger fractions confirm the time-scheme is converging. 

Although conceptually simple, the Lagrangian Monte Carlo algorithm requires careful implementation to efficiently support an extremely high number of realizations. Given the algorithm's inherently parallel structure, GPU architectures are the natural choice. To this end, we have developed a CUDA+MPI implementation that minimizes communication -- the real bottleneck in reduction-heavy algorithms -- by hierarchically distributing realizations and reductions across nodes, GPU devices, GPU blocks, warps, and, finally, batches in each thread. The Philox\_4x32\_10 (pseudo)random number generator from the cuRAND library is used.
This implementation allowed us to reach $10^{13}$ realizations with a computational cost of $\sim 7$ GPU-hours.

\bibliography{main}
\end{document}